%% file: main.tex
\documentclass[10pt, paper=a4, UKenglish]{article}
\usepackage{graphicx}
\usepackage{amsmath}
\usepackage{tikz}
\usetikzlibrary{shapes,arrows}
\usetikzlibrary{positioning}
\usepackage{authblk}

\def\Title#1{\begin{center} {\Large #1 } \end{center}}
\def\Author#1{\begin{center}{ \sc #1} \end{center}}

\newcommand\pubblock{\rightline{\begin{tabular}{l} Proceedings of the CTD 2022\\ \pubnumber\\
			\pubdate  \end{tabular}}}

\newcommand{\ttbar}{\ensuremath{\text{t}\bar{\text{t}}}}
\newcommand{\ptvecmiss}{\ensuremath{{\vec p}_{\mathrm{T}}^{\kern1pt\text{miss}}}}

\newcommand{\pt}{\ensuremath{\text{P}_\text{T}}}

\newcommand{\dxy}{\ensuremath{\Delta_\text{xy}}}
\tikzstyle{block} = [rectangle, draw, fill=blue!20, 
text width=5em, text centered, rounded corners, minimum height=2em]
\tikzstyle{line} = [draw, -latex']
\tikzstyle{cloud} = [draw, ellipse,fill=red!20, node distance=3cm,
minimum height=2em]

\newenvironment{Abstract}{\begin{quotation} \begin{center} 
			\large ABSTRACT \end{center}\bigskip 
		\begin{center}\begin{large}}{\end{large}\end{center} \end{quotation}}

\newenvironment{Presented}{\begin{quotation} \begin{center} 
			PRESENTED AT\end{center}\bigskip 
		\begin{center}\begin{large}}{\end{large}\end{center} \end{quotation}}

\def\Acknowledgements{\bigskip  \bigskip \begin{center} \begin{large}
			\bf ACKNOWLEDGEMENTS \end{large}\end{center}}

\input econfmacros.tex

\usepackage{color}
\usepackage{lineno}
\usepackage{hyperref}
\usepackage{subcaption}
\usepackage[sorting=none]{biblatex}
\newcommand\pubnumber{PROC-CTD2022-10}

\newcommand\pubdate{\today}

\newcommand{\conference}{Connecting the Dots Workshop (CTD 2022)\\
	May 31 - June 2, 2022}

\usepackage{fancyhdr}
\definecolor{mygrey}{RGB}{105,105,105}
\begin{document}
	\large
	\begin{titlepage}
		\pubblock
		\vfill
		\Title{Line Segment Tracking in the HL-LHC}
		\vfill
	    \Author{Gavin Niendorf, Tres Reid, Peter Wittich (Cornell University)}
		\Author{Peter Elmer, Bei Wang (Princeton University)}
		\Author{Philip Chang, Yanxi Gu, Vyacheslav Krutelyov, Balaji Venkat Sathia Narayanan, Matevž Tadel, Emmanouil Vourliotis, Avi Yagil (UC San Diego)} 
		\vfill
		\begin{Abstract}
			The major challenge posed by the high instantaneous luminosity in the High Luminosity LHC (HL-LHC) motivates efficient and fast reconstruction of charged particle tracks in a high pile-up environment. While there have been efforts to use modern techniques like vectorization to improve the existing classic Kalman Filter based reconstruction algorithms, Line Segment Tracking takes a fundamentally different approach by doing a bottom-up reconstruction of tracks.  Small track stubs from adjoining detector regions are constructed, and then these track stubs that are consistent with typical track trajectories are successively linked. Since the production of these track stubs is localized, they can be made in parallel, which lends way into using architectures like GPUs and multi-CPUs to take advantage of the parallelism. The algorithm is implemented in the context of the CMS Phase-2 Tracker and runs on NVIDIA Tesla V100 GPUs. Good physics and timing performance has been obtained, and stepping stones for the future are elaborated. 
		\end{Abstract}
		
		\vfill
		
		\begin{Presented}
			\conference
		\end{Presented}
		\vfill
	\end{titlepage}
	\def\thefootnote{\fnsymbol{footnote}}
	\setcounter{footnote}{0}
	%
	
	\normalsize 
	
	\section{Introduction}
\begin{figure}[!htb]
		\centering
		\includegraphics[width = 0.5\textwidth]{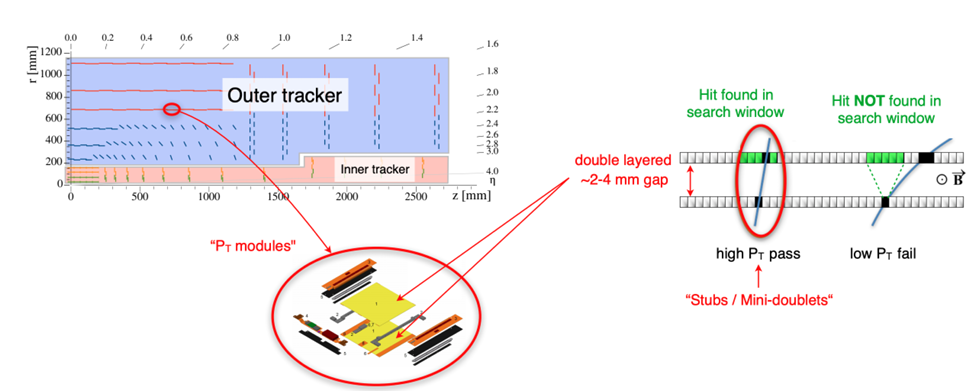}
		\caption{Outer tracker geometry and module structure in CMS at the HL-LHC~\cite{CERN-LHCC-2017-009}}
		\label{fig:hllhctracker}
	\end{figure}

	The reconstruction of charged particle tracks is at the heart of object reconstruction in the  experiments at the Large Hadron Collider. The tracker provides accurate estimates of momentum and energy of the decay particles and hence plays a crucial role in the reconstruction of parent particles. Algorithms such as vertex and jet reconstruction, and calculation of the missing transverse momentum and b-jet tagging rely crucially on tracking.
	\paragraph{}
	Charged particle track reconstruction is the most expensive and the most time consuming step in the object reconstruction pipeline. Since this process is combinatorial in nature, it scales exponentially with the multiplicity of hits in the detector. With increased pp collisions in the HL-LHC~\cite{HLLHC}, the pile-up will increase around 4-5 times compared to its existing value in the LHC. This implies that tracking times will increase by more than an order of magnitude in the HL-LHC if we stay with the same tracker and algorithms used in the CMS Experiment during Run 2~\cite{Cerati:1966040}. This implies the existing iterative tracking algorithms will be too slow to be able to efficiently reconstruct tracks in a timely manner in the HL-LHC. In addition, computational performance of single thread processors is plateauing while computational demands are increasing. However, the current era is also marked by the advent of multi-threaded and multi-core CPUs, and innovations in the co-processor sphere like Graphical Processing Units (GPUs) and Field Programmable Gate Arrays (FPGAs) which are able to provide massive increases in computational processing speed, and hence are expected to dominate High Performance Computing (HPC) workflows in the near future. This motivates a re-design of track reconstruction algorithms to take advantage of these new architectures and break the computational barrier in the near future. 
	\paragraph{}
	In this work, we present a bottom-up localized track reconstruction algorithm for the CMS outer tracker (Figure~\ref{fig:hllhctracker}) called Line Segment Tracking, wherein charged particles are grouped together to reconstruct entire tracks from the ground up. An advantage of this algorithm is that it is readily parallelizable, since only localized information is required at each step to reconstruct track constituents.
	The outer tracker (Figure~\ref{fig:hllhctracker}) is composed of bi-layer \pt~ modules (Figure~\ref{fig:hllhctracker} right). A \pt~module has two silicon layers that are $\approx$ 2 to 4mm apart. These closely spaced modules ensure hits can be correlated to get preliminary estimates of the transverse momentum, using which track stubs made up of hit pairs consistent with a track hypothesis can be selected. These track stubs can be used as fundamental building blocks of any tracking algorithm. The usage of the track stubs ensure that the effective number of initial seeds is reduced, thereby reducing combinatorics without any loss of efficiency.
	\paragraph{}
	Line Segment Tracking was originally inspired by the XFT algorithm used at the CDF Experiment at the Tevatron, Fermilab~\cite{XFT}. A previous version of this algorithm was introduced in the 2020 edition of Connecting the Dots~\cite{chang2020parallelizable}.  A similar algorithm was used in the context of grouped layer tracker layout design~\cite{krutelyov2016} which showed promising results on the combinatorics reduction front. 
	\section{Line Segment Tracking in the HL-LHC CMS Outer Tracker}
	\label{sec:lst}
	\subsection{Overview}
	Line Segment Tracking (LST) is a parallelized track reconstruction algorithm that does a bottom-up reconstruction of tracks by creating track objects from correlated hit pairs (stubs). These short objects are then linked to create longer objects with a greater number of associated hits. This process is repeated until entire tracks are reconstructed. Since the reconstruction of the higher order objects only depends on lower order objects in the immediate neighbourhood, the reconstruction is highly local and hence can be parallelized effectively. The hits are first clustered into Mini-doublets, which are comprised of two hits. Mini-doublets then link up to create segments (four hits). Segments then link up to create Triplets (6 hits), which can then be linked to create Quintuplets (10 hits) and so on. Various physics and geometrical selections are applied at each linking level to reduce combinatorics and ensure only legitimate track candidates are created.
	\paragraph{}
	In addition, tracking seeds from the inner pixel tracker are incorporated as segments, which when linked with a Triplet or a Quintuplet can create a Pixel Triplet or a Pixel Quintuplet. The addition of the pixel track seeds helps in reducing the combinatorial fakes and essentially provides a link between the inner and outer trackers.
	\subsection{Mini-doublets and Segments}
		\label{subsec:mdseg}
\begin{figure}[!htb]
	 	\centering
	 	\includegraphics[width = 0.3\textwidth]{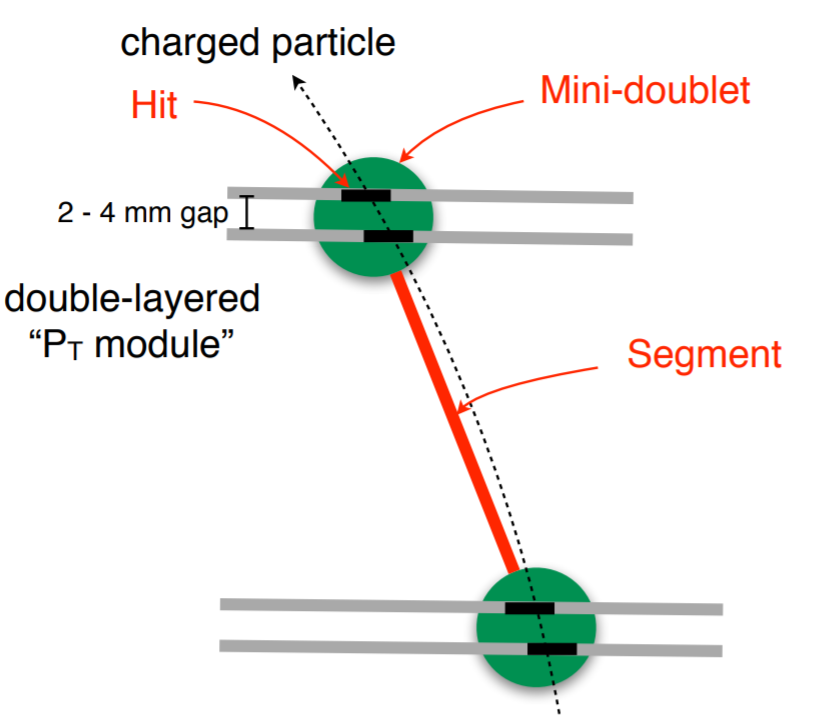}
	 	\caption{Mini-doublets (Green dots) linked using module maps to create segments}
	 	\label{fig:mdseg}
	 \end{figure}

	 The track stub created in a bi-layer module consistent with the \pt\, threshold is called a Mini-doublet. These are denoted by the green dot in Figure~\ref{fig:mdseg}. These are the fundamental building blocks of our algorithm. Since we only have localized information at this juncture, we only rely on the slope information consistent with a particle coming from the origin, and account for additional error terms which take care of multiple scattering and uncertainties in the luminous region.
	 \paragraph{}
	 Line Segments are produced by linking two mini-doublets from modules across different layers (Figure~\ref{fig:mdseg}). A ``segment candidate'' is formed from two mini-doublets in different layers and then physics selections are applied to check if this can be a legitimate segment from a true track. To reduce our search window for the number of possible connections, we derive a ``module map'' which is a lookup table that details the list of possible module connections from a given module. This was derived from first principles by computing the possible track trajectories as helices that can start from the edges of a given module that are  compatible with the \pt threshold (in our case, 0.8 GeV), and then listing all the modules in the adjacent layer that intersect these helices. This map was then verified using simulated samples for correctness. A first principles approach like this helps us take all possible edge cases into account, which would be quite difficult if we relied only on simulations to compute the module map. 

	\subsection{Triplets and Quintuplets}
\begin{figure}[!htb]
    \centering
        \begin{subfigure}{0.48\textwidth}
            \includegraphics[width = \textwidth]{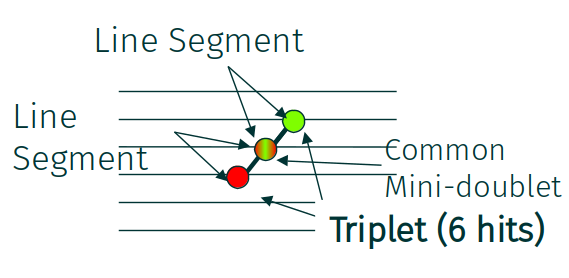}
            \caption{Triplet formed from two line segments}
            \label{fig:triplets}
        \end{subfigure}
        \begin{subfigure}{0.48\textwidth}
            \includegraphics[width = \textwidth]{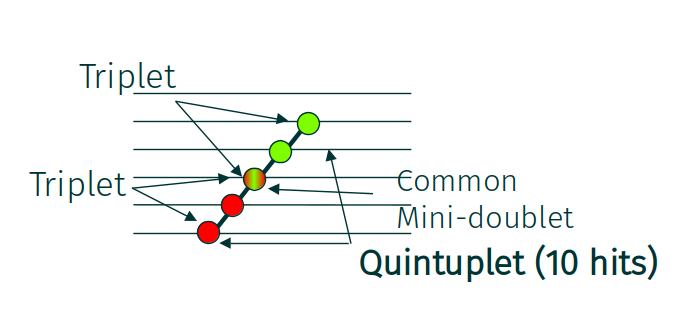}
            \caption{Quintuplet formed from two Triplets}
            \label{fig:quintuplets}
        \end{subfigure}
        \caption{Triplet and Quintuplet}
        \label{fig:T3T5}
    \end{figure}

	Two line segments having a common mini-doublet can be linked to form a Triplet (also called a T3), as shown in Figure~\ref{fig:triplets}. The requirement of a common mini-doublet greatly reduces the combinatorics. Other selections here include the $\chi^2$ obtained from fitting a straight line to the three ``anchor hits'' of the mini-doublet in the r-z plane, the anchor hit being the more-precise hit on the pixel side for a PS module, and the hit on the lower side of a 2S module.
	\paragraph{}
	Similarly, two triplets with a common mini-doublet can be joined to form a Quintuplet (also called a T5), as shown in Figure~\ref{fig:quintuplets}. In addition to applying a straight line pointing criterion similar to the triplet case, we also require that the circles formed from the anchor hits of the inner and outer triplets have radii close to each other. Such circles are trivially created from the three points of a triplet. This is followed by a global circle fit using all the five anchor hits and a selection based on the $\chi^2$ of the resulting hit. The $\chi^2$ cuts are derived empirically for different categories of quintuplets and provide good reduction of the spurious linkages. 
    \subsection{Linking the pixel and outer trackers - Pixel Triplets and Pixel Quintuplets}
\begin{figure}[!htb]
        \centering
        \begin{subfigure}{0.48\textwidth}
            \includegraphics[width = \textwidth]{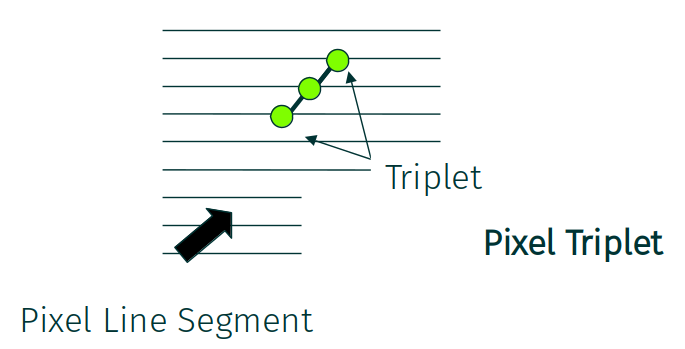}
            \caption{Pixel Triplet}
            \label{fig:pixeltriplets}
        \end{subfigure}
        \begin{subfigure}{0.48\textwidth}
            \includegraphics[width = \textwidth]{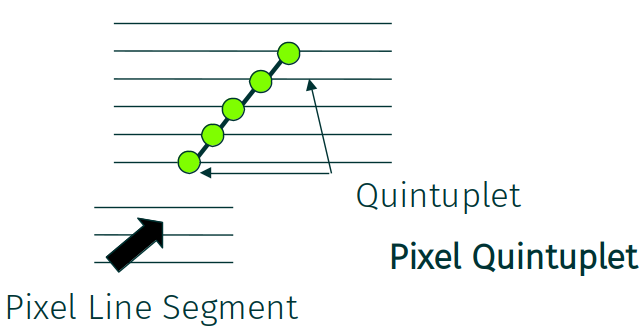}
            \caption{Pixel Quintuplet}
            \label{fig:pixelquintuplets}
        \end{subfigure}
        \caption{Pixel linked objects - Pixel Triplet and Pixel Quintuplet}
        \label{fig:pT3pT5}
    \end{figure}

    Pixel track seeds produced in the inner tracker are incorporated in LST as line segments, called Pixel Line Segments (pLS). The outer tracker objects can be linked to these pixel line segments to produce objects such as Pixel Triplets and Pixel Quintuplets (Figure~\ref{fig:pT3pT5}). Selections similar to that used for the Quintuplets are used here, the difference being that the straight lines and circles in the r-z and x-y plane, respectively, are not fit but rather the estimates from the pixel detector are used and the consistency of the outer hits to this hypothesis is checked. The incorporation of the pixel line segments ensure that only those tracks from the interaction vertex that are also consistent with the inner pixel are ultimately incorporated, thereby reducing combinatorial fakes even further.
    
    \subsection{Cross-cleaning and Track Candidates}
    The final track candidate collection consists of pixel quintuplets, pixel triplets, Quintuplets and pixel line segments, added in that sequence.  The Pixel Quintuplets are the cleanest objects that also span almost the entire length of the outer detector. So these will serve as the main component of our final track candidates. To ensure track duplicates between different objects are not added to the final collection, we ``cross-clean'' the objects. So when Pixel Triplets are added to the final collection, only those Pixel Triplets that are not near a Pixel Quintuplet (in $\eta$-$\phi$ space) are added. This ensures that the Pixel Triplets added are those that exclusively correspond to those tracks which could not produce a Pixel Quintuplet possibly due to a missing mini-doublet and/or hit somewhere that resulted in the underlying objects not being produced. Similarly, the Quintuplets after undergoing cross-cleaning correspond mainly to the displaced tracks, since they will not have any pixel track seeds (this can be seen in the transverse impact parameter, \dxy, efficiency distributions in Figure~\ref{fig:effvsdxypu200}). Finally, the Pixel Line Segments that follow a strict set of criteria (they need to not be a part of any other object, they should be appreciably far away in $\eta-\phi$ space with the other objects etc) provide coverage mainly in the forward region (this can be seen in the $\eta$ efficiency distributions in Figure~\ref{fig:effvsetastack}). 
    \subsubsection{Track Extensions and Final Track Collection}
    A subset (approximately half) of the track candidates can be linked with triplets in the outermost regions of the tracker to create longer tracks that can span the entire length of the outer tracker. These extended tracks along with those track candidates that cannot be extended, make up the final collection of track candidates by the algorithm.
    
    \section{Physics performance}
	\label{sec:physics}
	To measure the physics performance and fine-tune the cuts, a dataset of the simulated detector response to  a \ttbar\, decay in a pile-up 200 environment was produced. A track candidate produced by Line Segment Tracking is said to be matched to a ``true'' (simulated) track if 75\% of the associated hits match. The efficiency is computed as the fraction of true tracks in a \pt\, or $\eta$ bin that have a matched track candidate. 
	\paragraph{}
	The fake rate is calculated as the fraction of reconstructed tracks that are fakes (not matched to true tracks), while the duplicate rate is calculated as the fraction of reconstructed tracks that are duplicates. When two or more reconstructed tracks match with a simulated track, all of them are considered duplicates of one another and get included in the computation of the duplicate rate.
	
	\subsection{Efficiency performance}
	\begin{figure}[!htb]
	    \centering
		\begin{subfigure}{0.4\textwidth}
			\includegraphics[width = \textwidth]{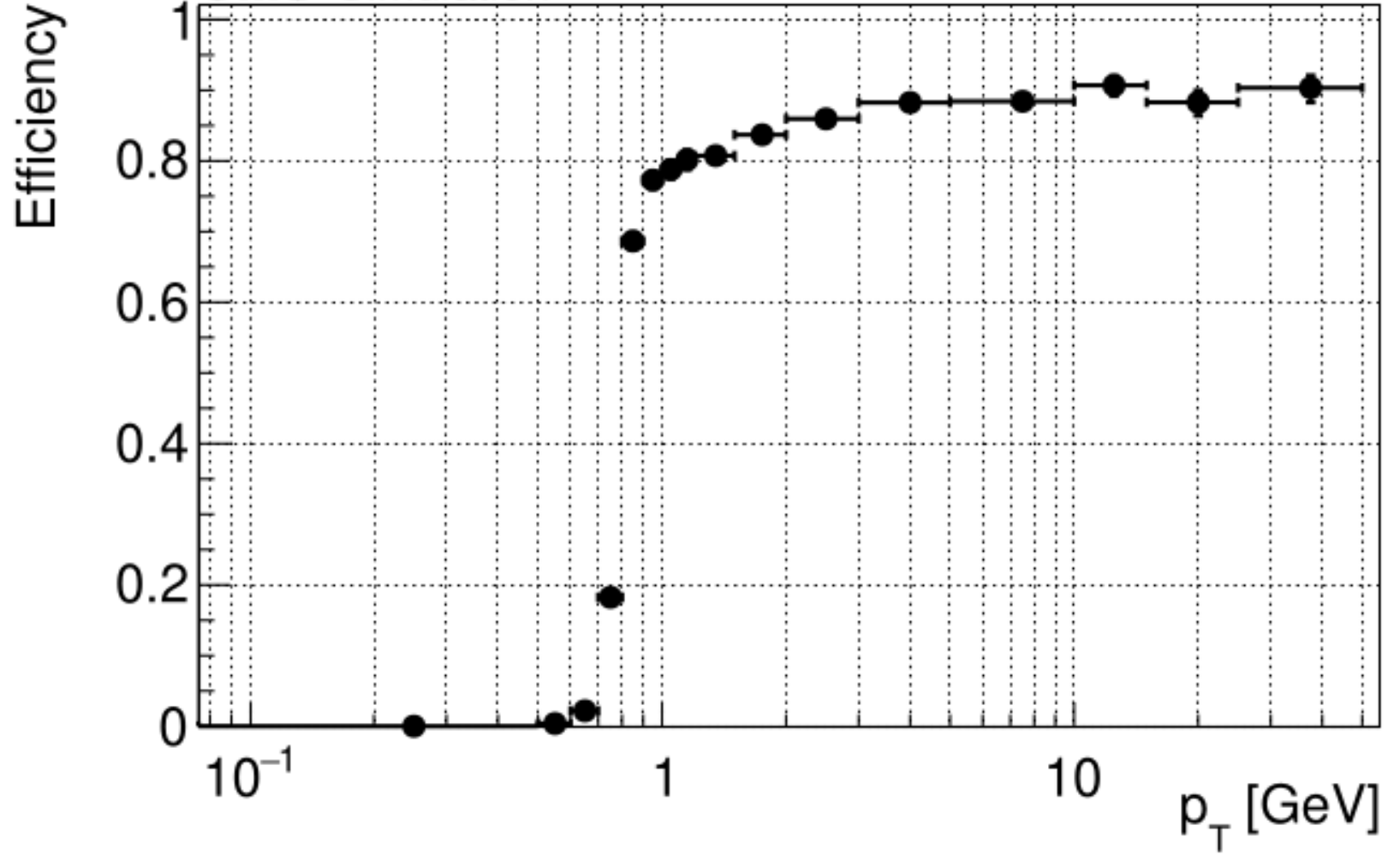}
			\caption{Efficiency vs~\pt}
			\label{fig:effvspt}
		\end{subfigure}
		\begin{subfigure}{0.4\textwidth}
			\includegraphics[width = \textwidth]{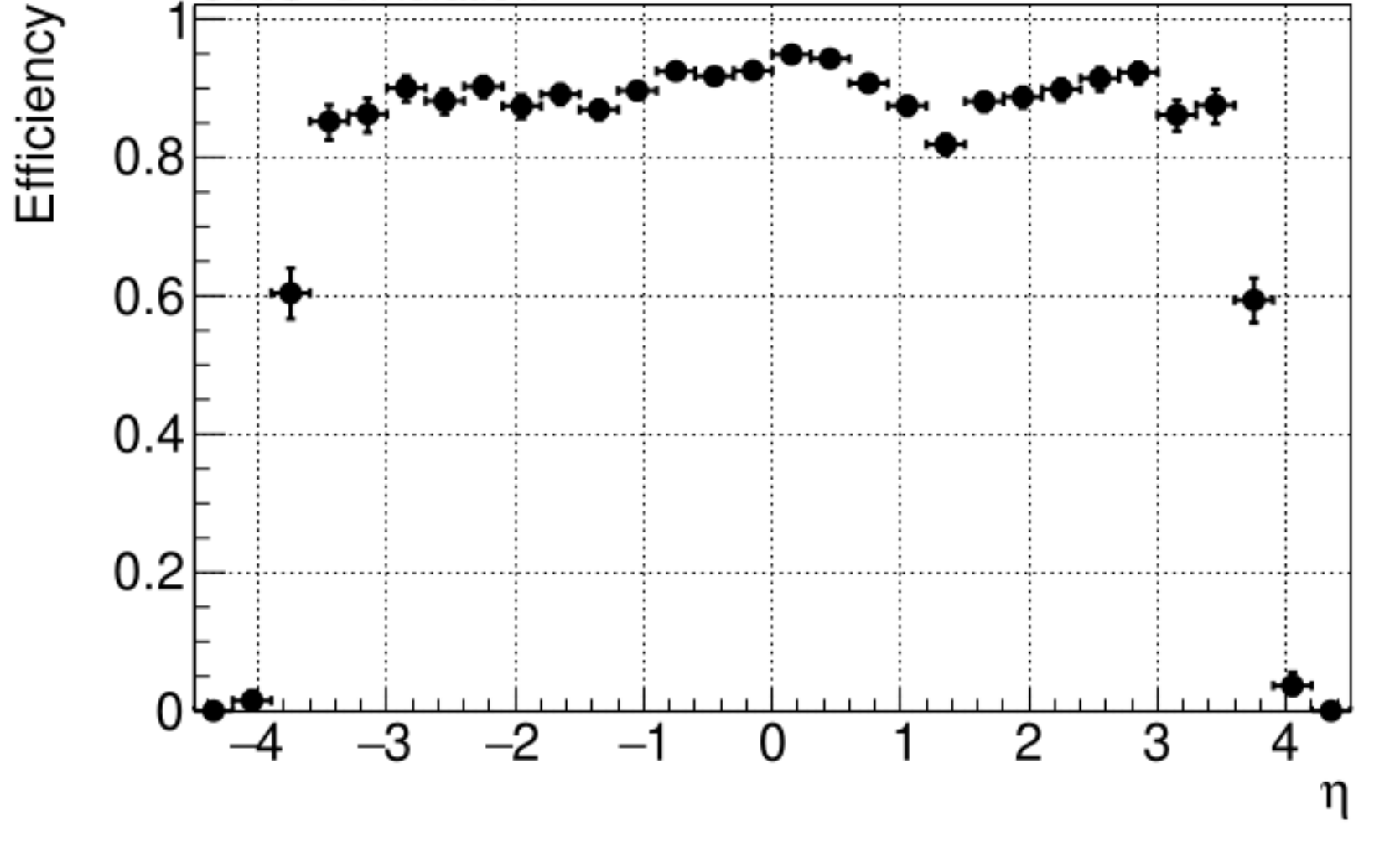}
			\caption{Efficiency vs~$\eta$}
			\label{fig:effvseta}
		\end{subfigure}
		\caption{LST efficiency on \ttbar~+ PU200 sample}
		\label{fig:effpu200}
		\end{figure}
\begin{figure}[!htb]
	   \centering
	\begin{subfigure}{0.4\textwidth}
		\includegraphics[width = \textwidth]{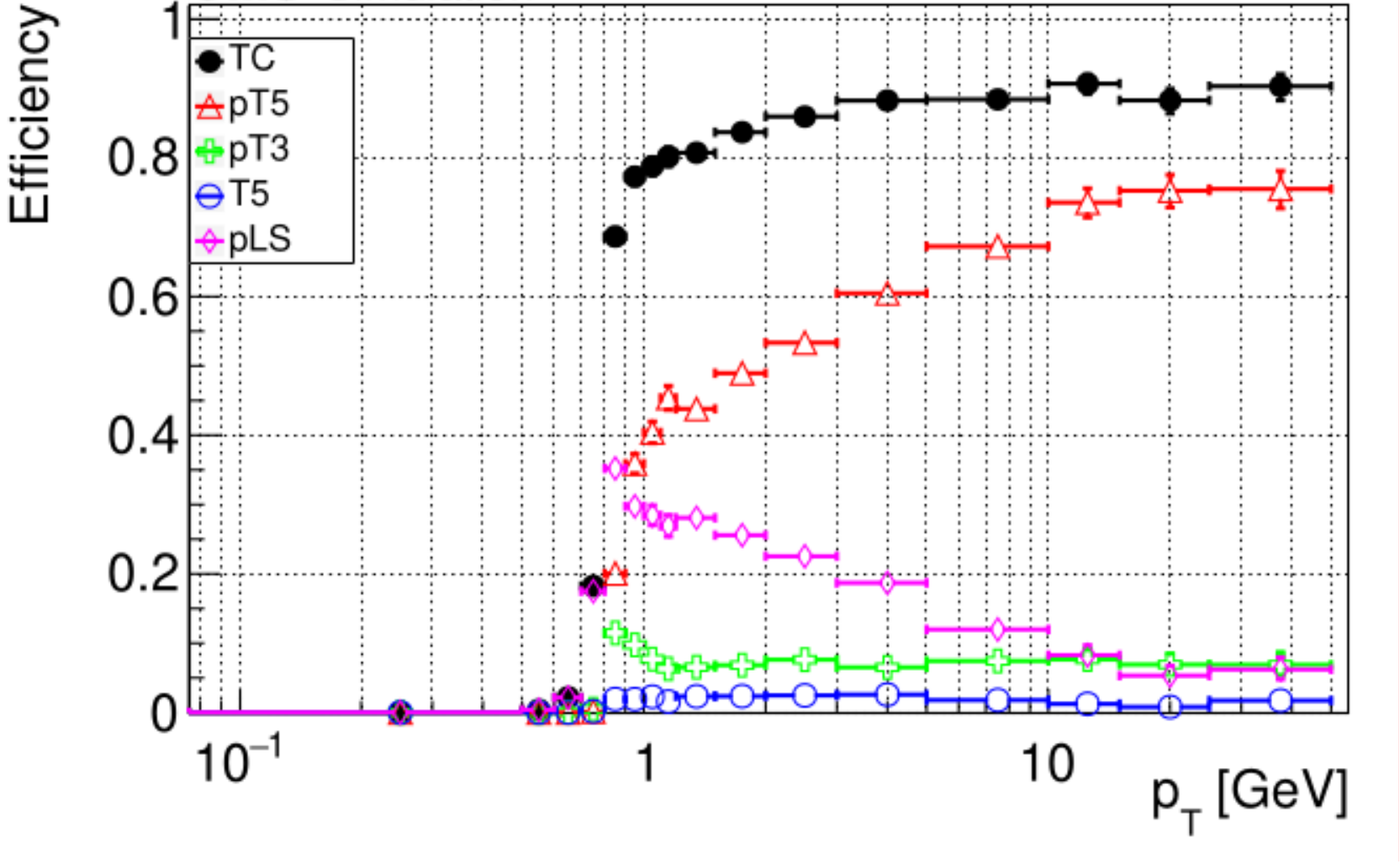}
		\caption{Efficiency vs~\pt}
		\label{fig:effvsptstack}
	\end{subfigure}
	\begin{subfigure}{0.4\textwidth}
		\includegraphics[width = \textwidth]{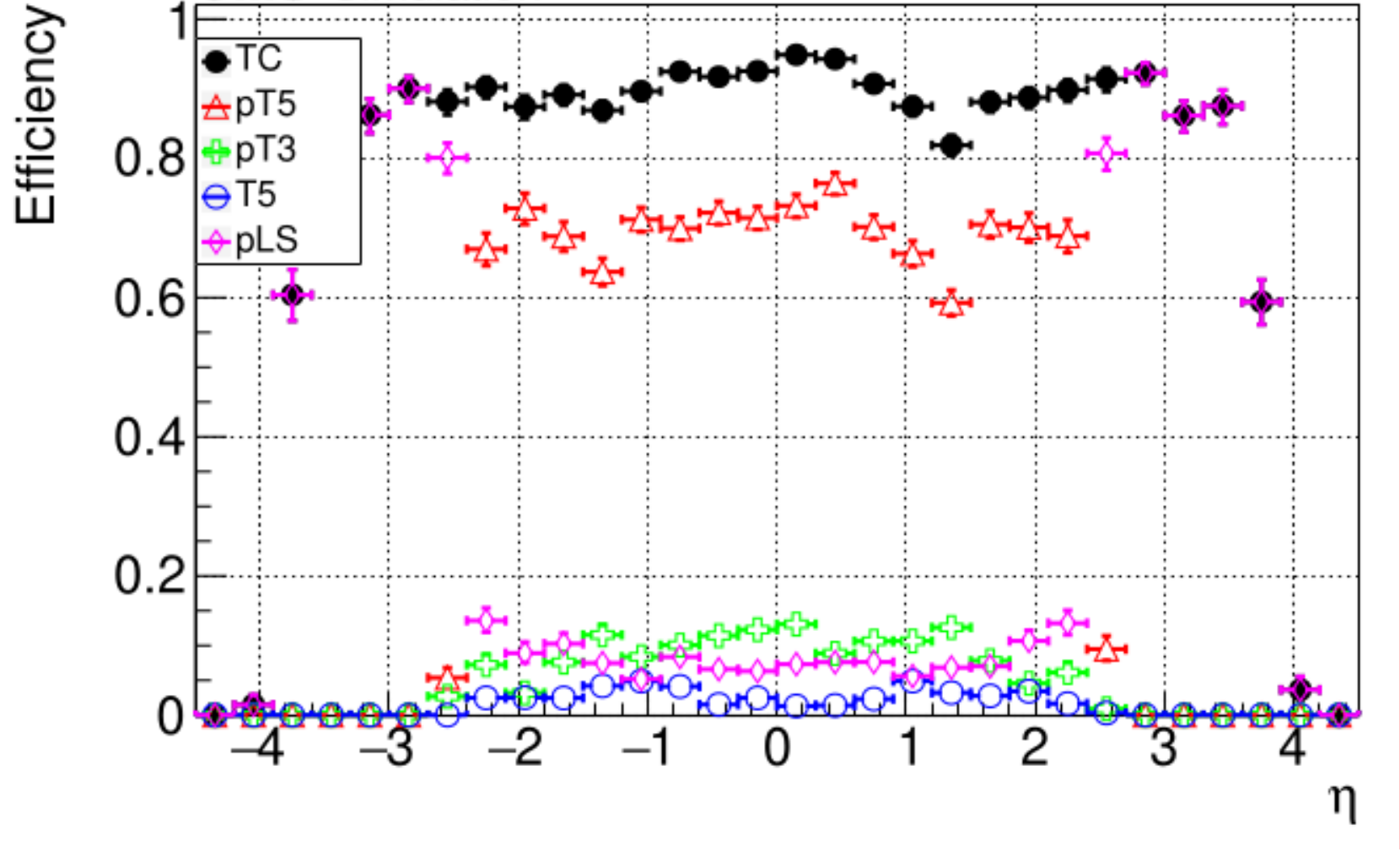}
		\caption{Efficiency vs~$\eta$}
		\label{fig:effvsetastack}
	\end{subfigure}
	\caption{LST efficiency on \ttbar~+ PU200 sample}
	\label{fig:effstackpu200}
\end{figure}
\begin{figure}[!htb]
	   \centering
	\begin{subfigure}{0.4\textwidth}
		\includegraphics[width = \textwidth]{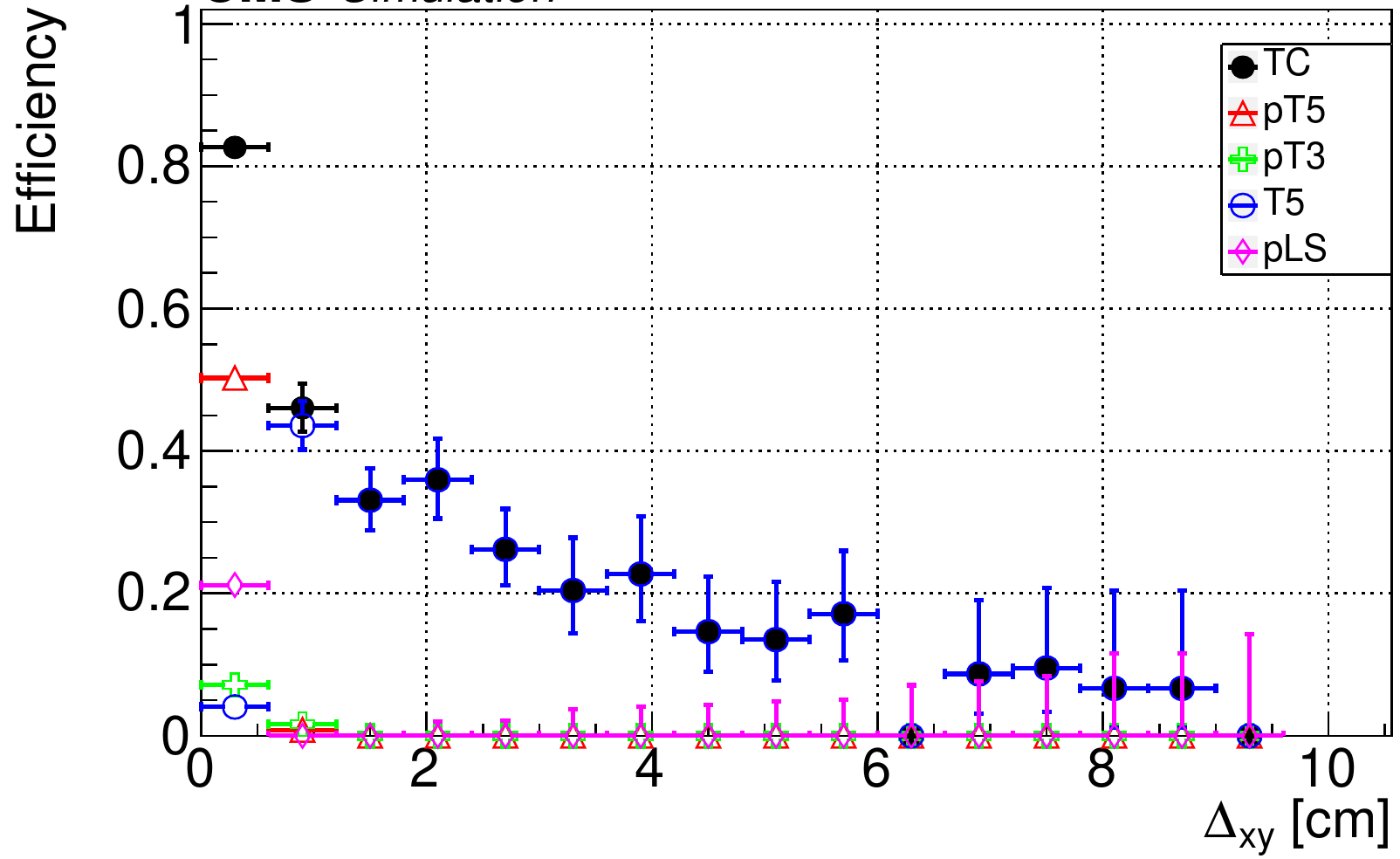}
		\caption{Efficiency vs~\dxy~in \ttbar~+ PU200 sample}
		\label{fig:effvsdxypu200}
	\end{subfigure}
	\begin{subfigure}{0.4\textwidth}
		\includegraphics[width = \textwidth]{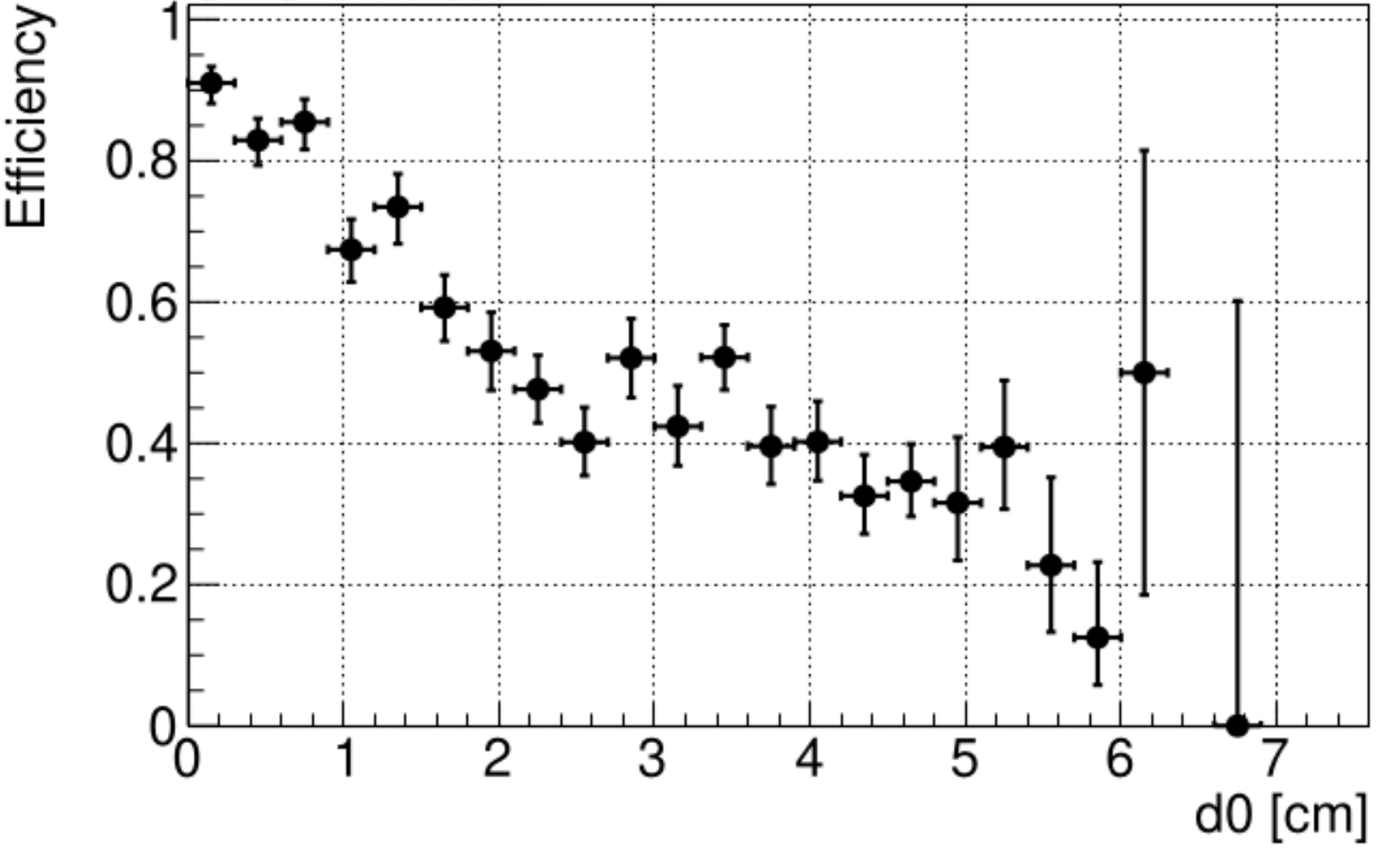}
		\caption{Efficiency vs~\dxy~in 5cm muon cube sample}
		\label{fig:effvsdxycube}
	\end{subfigure}
	\caption{LST efficiency with displaced tracks - PU200 and muon cube}
	\label{fig:effdxy}
\end{figure}

	The track candidate reconstruction efficiency distributions as a function of \pt\, and $\eta$ are shown in Figures~\ref{fig:effvspt} and \ref{fig:effvseta} as a function of \pt\, and $\eta$, respectively. The turn-on region is at 0.8 GeV, and we see good reconstruction efficiency above that, comparable with the existing Combinatorial Kalman Filter Based tracking algorithm for the CMS~\cite[Figure 10.2]{tdr}. The $\eta$ distributions are computed with an implicit \pt~cut of 0.9 GeV. The contributions to the track candidate efficiency plots broken up into individual components are shown in Figure~\ref{fig:effstackpu200}. The Pixel Quintuplets (pT5) contribute the highest to the efficiency, while the Pixel Line Segments (pLS) not linked to an outer tracker object contribute in the forward ($|\eta| > 2$) regions. 
    \paragraph{}
    The displaced track reconstruction performance is measured by computing the efficiency as the function of the x-y distance from the interaction point ($\Delta_\text{xy}$). In addition to checking the physics performance on the \ttbar + Pile-up 200 sample (Figure~\ref{fig:effvsdxypu200}) in which the Quintuplets (T5) contribute the most to displaced track reconstruction (non-zero $\Delta_\text{xy}$), we also verified the physics performance on a sample of displaced muon tracks originating from a point displaced from the origin in a 5cm cube (Figure~\ref{fig:effvsdxycube}). Good reconstruction efficiency has been achieved, given the fact that no extra steps in the algorithm are dedicated specifically to reconstructing displaced tracks. These promising results show that this can be improved further.

\subsection{Fake and duplicate rates performance}
\begin{figure}[!htb]
			\centering
		\begin{subfigure}{0.4\textwidth}
			\includegraphics[width = \textwidth]{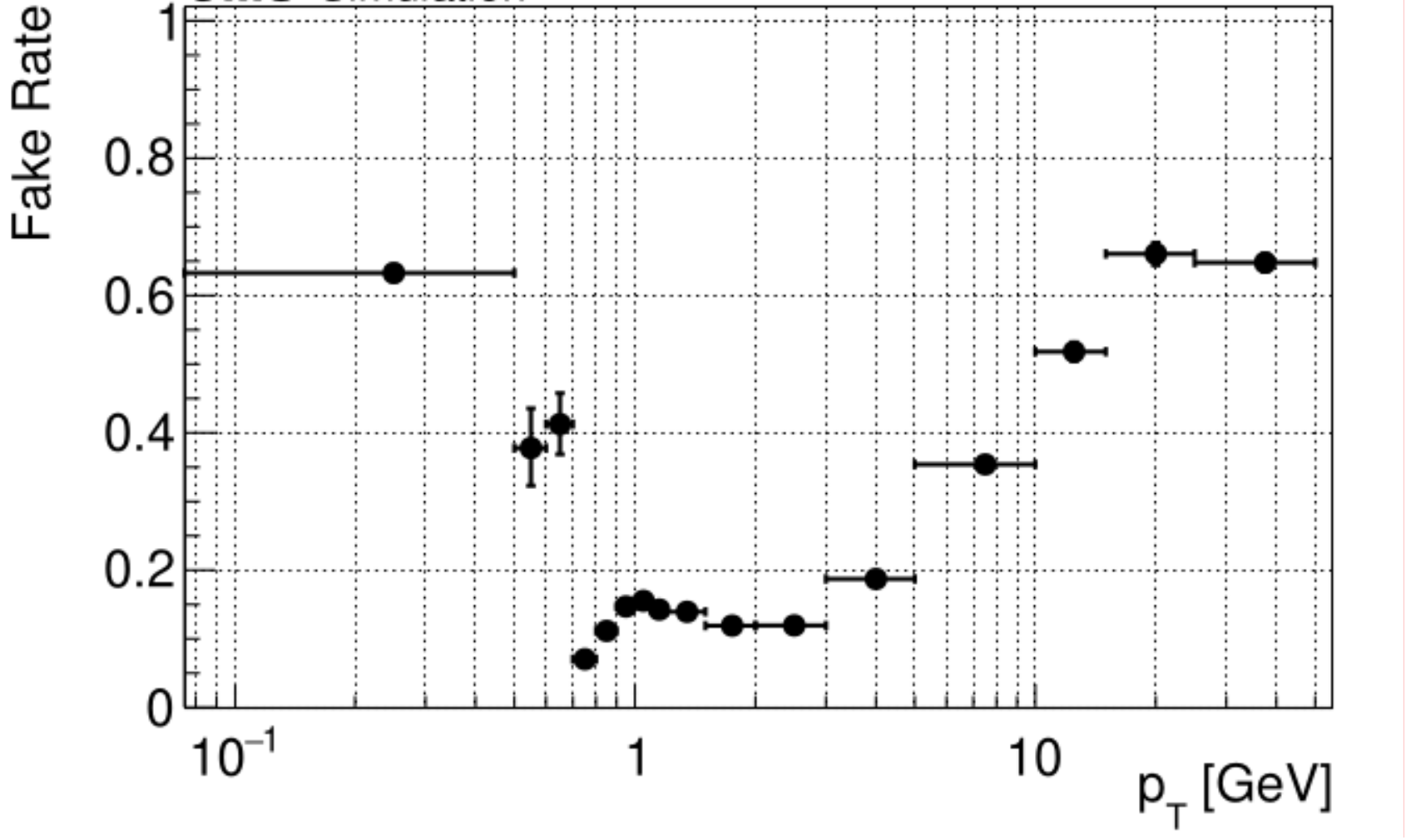}
			\caption{Fake rate vs~\pt}
			\label{fig:fakevspt}
		\end{subfigure}
		\begin{subfigure}{0.4\textwidth}
			\includegraphics[width = \textwidth]{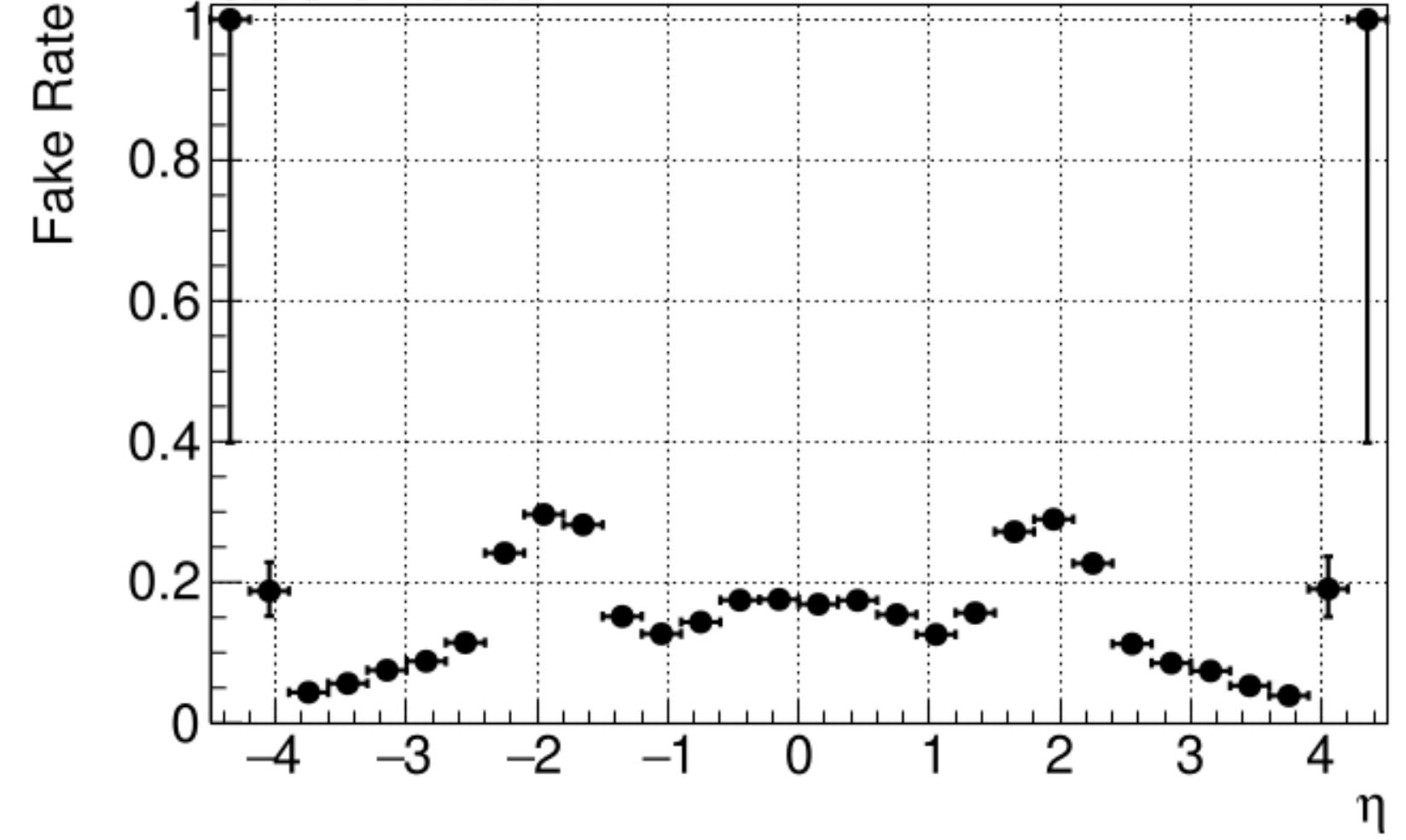}
			\caption{Fake rate vs~$\eta$}
			\label{fig:fakevseta}
		\end{subfigure}
		\caption{LST fake rates on \ttbar~+ PU200 sample}
		\label{fig:fakepu200}
	\end{figure}

\begin{figure}[!htb]
	\centering
	\begin{subfigure}{0.4\textwidth}
		\includegraphics[width = \textwidth]{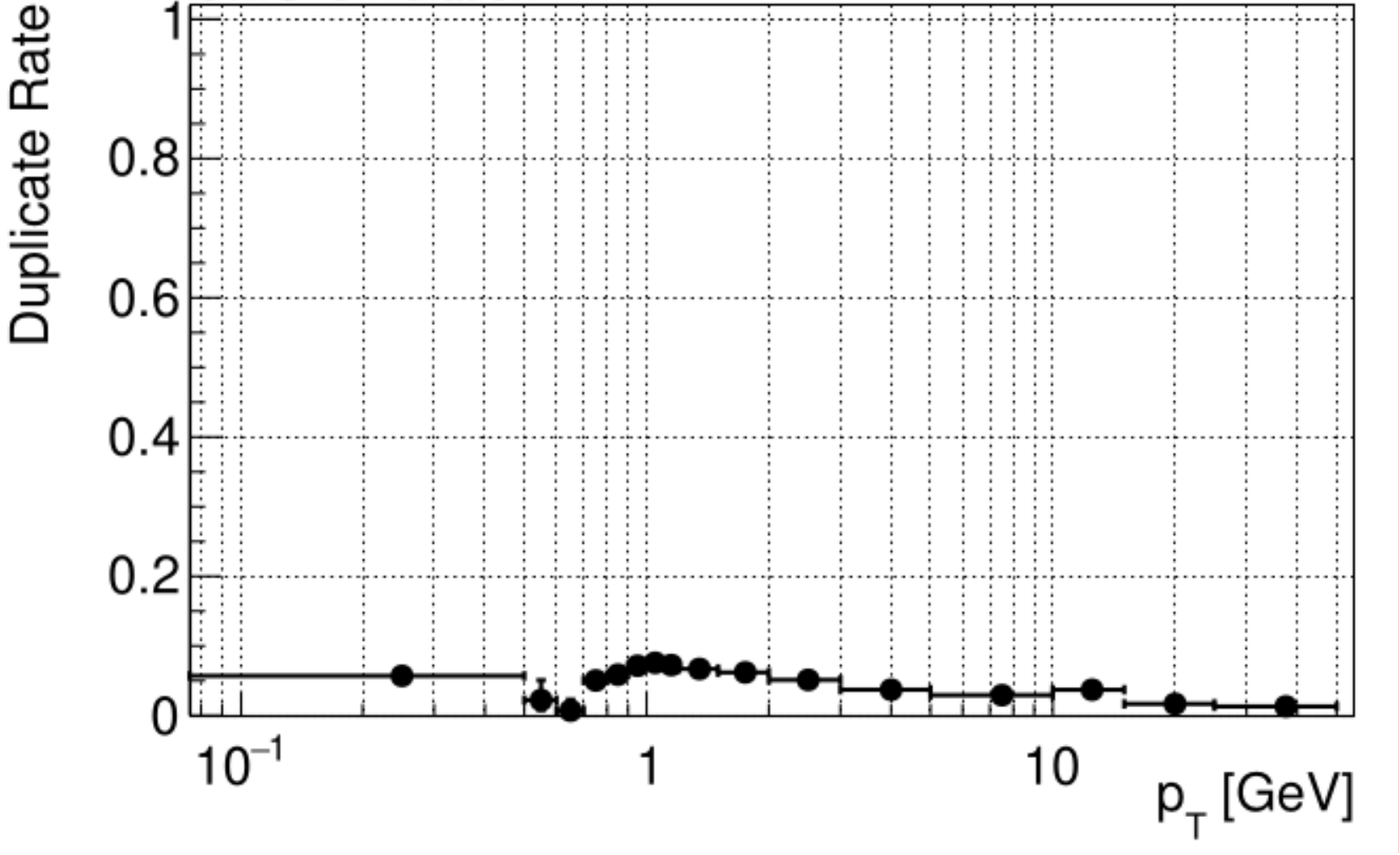}
		\caption{Duplicate rate vs~\pt}
		\label{fig:duplvspt}
	\end{subfigure}
	\begin{subfigure}{0.4\textwidth}
		\includegraphics[width = \textwidth]{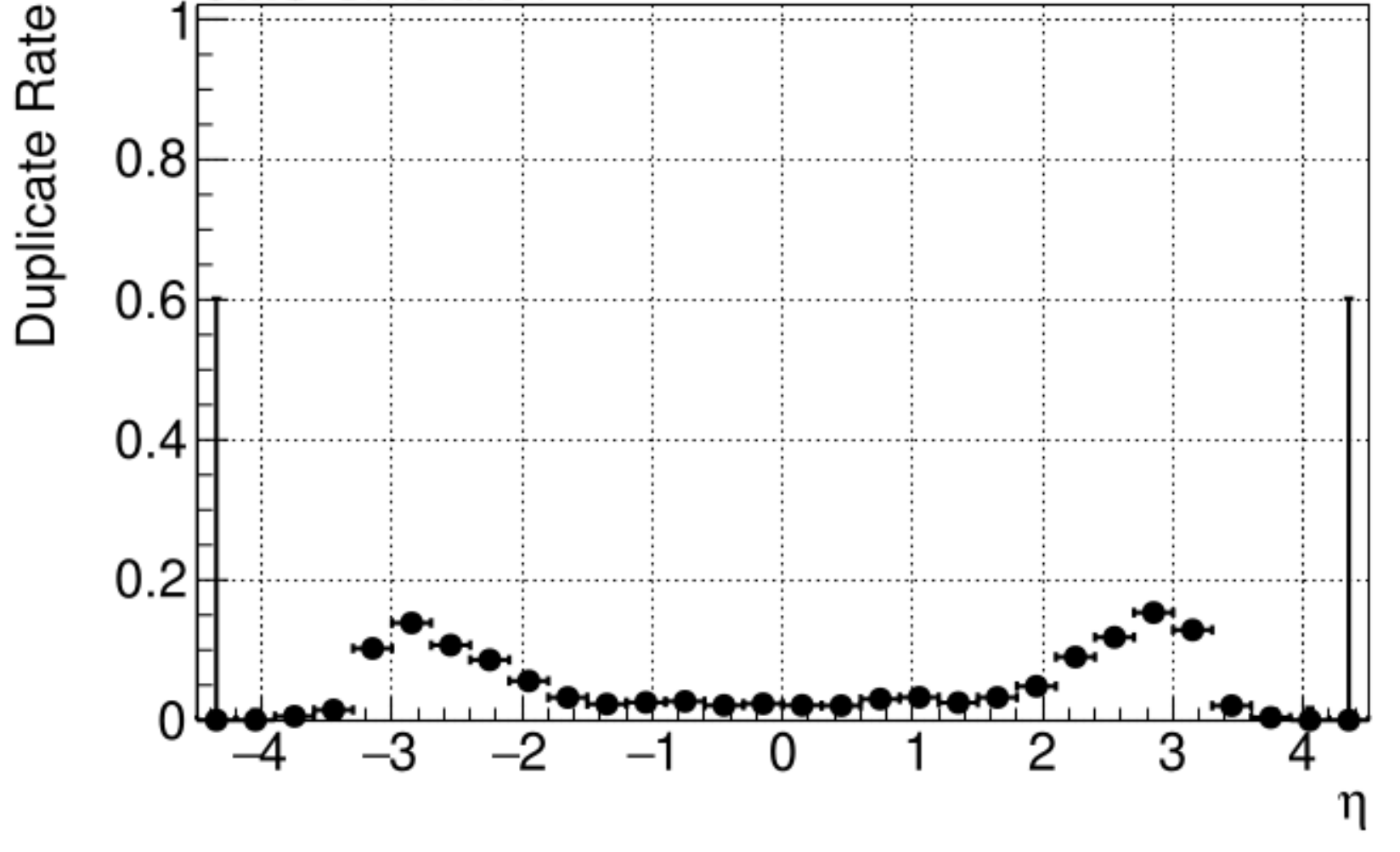}
		\caption{Duplicate rate vs~$\eta$}
		\label{fig:duplvseta}
	\end{subfigure}
	\caption{LST duplicate rates on \ttbar~+ PU200 sample}
	\label{fig:duplpu200}
\end{figure}

Figures~\ref{fig:fakevspt} and \ref{fig:fakevseta} show the distribution of fake rates as a function of \pt\, and $\eta$. The Pixel Line Segments (pLS) contribute the highest, especially in the forward region ($|\eta| > 2.5$), with almost all the fakes attributable to them in the forward region. These fake rates can be reduced further with fine-tuning of selection parameters and a full fit of hit patterns.
\paragraph{}
Since efficient cross-cleaning is done, the duplicate rates (Figures~\ref{fig:duplvspt} and \ref{fig:duplvseta}) are quite low. A low duplicate rate means that the final collection of tracks is compact and repeated reconstruction is reduced. 

	\section{Line Segment Tracking on the GPU}
	\label{sec:gpu}
	\subsection{GPUs overview}
\begin{figure}
	\centering
	  \includegraphics[width = 0.4\textwidth]{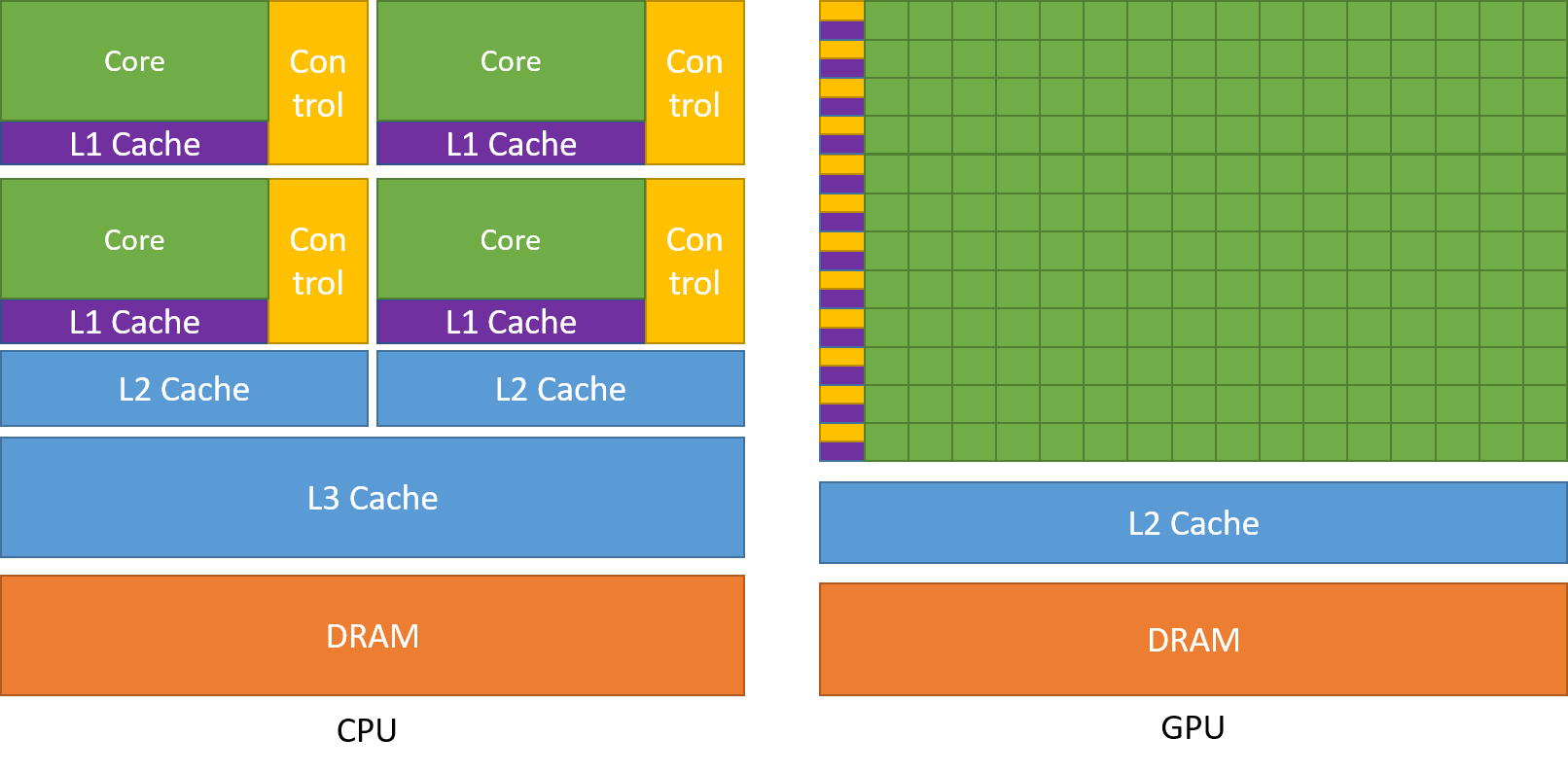}
	  \caption{CPUs vs GPUs~\cite{cuda} }
	  \label{fig:cpugpu}
	\end{figure}

	Graphical Processing Units (GPUs) (Figure~\ref{fig:cpugpu}) have numerous compute cores compared to CPUs. However they compromise on caches and data transport. In addition, GPUs also tend to have lower clock speed (approx 1 GHz), compared to CPUs (2-5 GHz).  If GPUs can be programmed such that the compute cores can do work on existing data while the caches wait for new data, tremendous speed-ups can be achieved. This technique is called latency hiding and is central to GPU programming.
	\paragraph{}
	CUDA (formerly an acronym for Compute Unified Device Architecture) is a GPU programming framework for Nvidia GPUs which provide high level functions to develop software that runs on GPUs. The Line Segment Tracking implementation for GPUs is written in CUDA and targets the recent generation of GPUs. 
	\subsection{GPU Implementation of Line Segment Tracking}
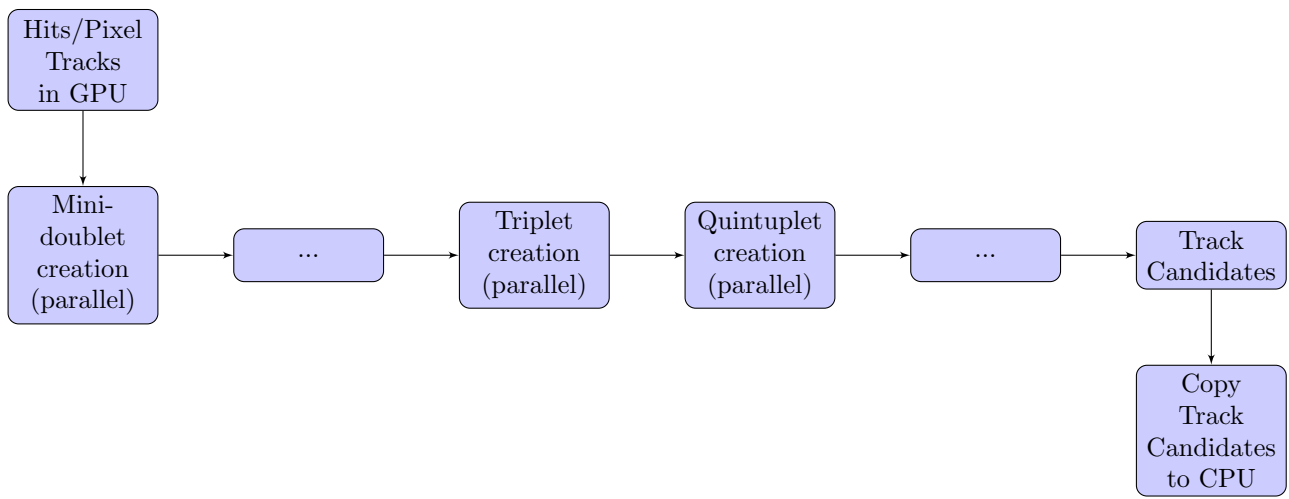
\begin{figure}[!htb]
	\centering
		\begin{tikzpicture}[node distance = 3cm ,auto]
			\node [block](hits) {Hits/Pixel Tracks in GPU};
			\node[block, below=1cm of hits](MD){Mini-doublet creation (parallel)};
			\node[block, right of=MD](dot1){...};
			\node[block, right of=dot1](T3){Triplet creation (parallel)};
			\node[block, right of=T3](T5){Quintuplet creation (parallel)};
			\node[block, right of=T5](dot2){...};
			\node[block, right of=dot2](TC){Track Candidates};
			\node[block, below=1cm of TC](copy){Copy Track Candidates to CPU};
			\path[line](hits) -- (MD);
			\path[line](MD) -- (dot1);
			\path[line](dot1) -- (T3);
			\path[line](T3) -- (T5);
			\path[line](T5) -- (dot2);
			\path[line](dot2) -- (TC);
			\path[line](TC) -- (copy);
        	\end{tikzpicture}
	    \caption{GPU implementation flowchart. Only the more important steps are explicitly shown, the others are represented by dots ($\dots$)}
	    \label{fig:gpuimplementation}
    \end{figure}

	Figure~\ref{fig:gpuimplementation} shows a visual representation of the GPU implementation. Each block, which corresponds to creating a particular class of track object, is parallelized. The entire flow, however, runs in sequence, since the algorithm requires that objects be hierarchically built upwards. Each of these object creation steps get their own kernel, in addition to a kernel that deals with pre-processing the hits and loading them into GPU memory. Multi-streaming and efficient memory management play an important role in improving performance.
    \subsubsection{Memory management}
    Since GPUs work on the Single Instruction Multi-thread (SIMT) principle, it is imperative that the data be stored in such a way that the compute cores get to use all the data that streams into the caches. Since GPUs are programmed such that multiple cores perform the same computational task in parallel on different data, cache hit rates and memory transfers from the device memory will be maximized when adjacent compute cores (that work on adjacent units of memory) get to use their respective data without any delays. This implies that the data need to be stored in a specific manner, commonly called the Structure of Arrays (SoA) data representation framework.
    \paragraph{}
    In the SoA data representation framework, the data corresponding to a particular quantity in memory is stored continuously for all objects in a single array. For example, if we want to store the \pt~ information for mini-doublets, we store this for all mini-doublets continuously in an array (as opposed to creating a class object for each mini-doublet, which will have the \pt\, value as one of its members). This would mean that when the \pt\, values are required for some computation, the compute cores will be readily able to access them with very high cache hit rates.
    \paragraph{}
    To reduce the huge overheads in memory allocation in the GPUs, we developed a caching allocator based on the CUB library developed by Nvidia~\cite{merrill2015cub}, which exponentially allocates memory based on the current demand, i.e., if 5MB of memory is requested, the caching allocator makes 8MB available, which means that the allocations for the next 3MB have zero overhead (when they happen). Since our framework has memory allocation and deletion spread all over, as opposed to all of it happening in one place in the beginning of the code, the caching allocator provides around 32\% improvement in compute time. 
    \subsubsection{Event level parallelization: Multi-streaming}
\begin{figure}[!htb]
    \centering
        \begin{subfigure}{0.9\textwidth}
        \includegraphics[width = 0.9\textwidth]{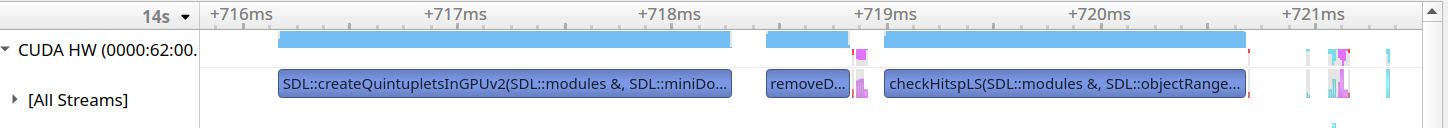}
        \caption{Single stream GPU workload timeline}
        \label{fig:nsys_1stream}
        \end{subfigure}
        \begin{subfigure}{0.9\textwidth}
        \includegraphics[width = 0.9\textwidth]{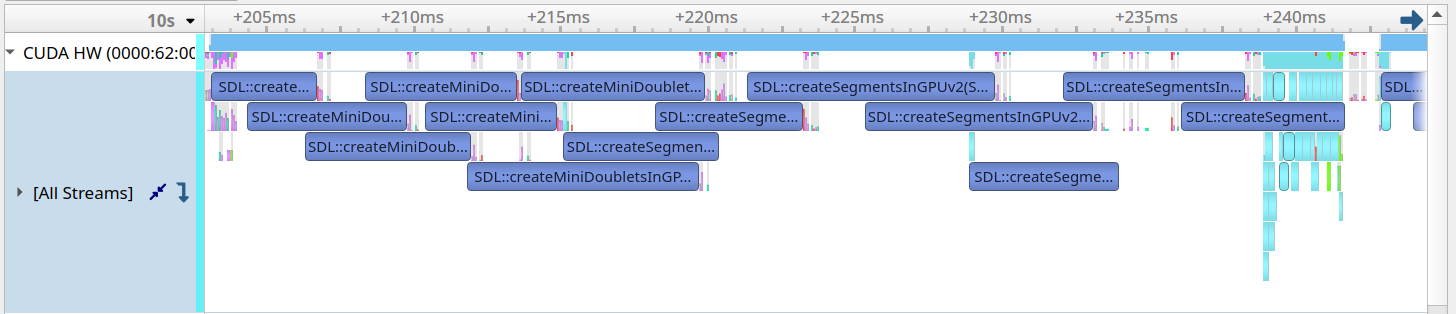}
        \caption{Eight streams GPU workload timeline}
        \label{fig:nsys_8streams}
        \end{subfigure}
        \caption{GPU timelines for the single stream and eight streams case}
        \label{fig:my_label}
    \end{figure}

    While classic event level parallelization tries to get multiple events, and their entire reconstruction processes to run in parallel, the sheer amount of compute cores required for object creation in the Line Segment Tracking case limits this application. However, the GPU downtimes can be used to do part of the work from multiple events in tandem. Figure~\ref{fig:nsys_1stream} shows the timeline of GPU usage when collision events get processed sequentially. The situation we have is that all the compute cores in the GPU are used for a period of time, followed by an interval of little to no work, which is when preparations for the next cycle of high intensity workload happens. Our variant of multi-streaming attaches events to streams, but the work gets distributed such that these downtimes are used for doing some work pertaining to a different stream. Figure~\ref{fig:nsys_8streams} shows how tasks from different streams (each row corresponds to a stream here) get tessellated. However, here are significant overlap regions where the GPU needs to work on two kernels, which means that compute cores get split between these kernels which result in a reduction in performance (and increase in time) of individual kernels, but the overall timing gets greatly improved.

	\subsection{Timing performance}
	\label{sec:timing}
	The complete timing performance, replete with individual object breakdowns when tracks from \ttbar + pile-up 200 events are reconstructed on a Tesla V100 GPU can be seen in Table~\ref{tab:timing}. Our best timing performance is 34 ms/event on a single stream, and 26 ms/event on eight streams. A caveat is that this timing performance does not take into account the extensive initial copy of hits from CPU to GPU, or the tracks from GPU to CPU. The timing performance is comparable to the latest results from the performance of the CMS Track Pattern recognition algorithms~\cite{dpnote}, which also reconstructs tracks with \pt~ above 0.8 GeV. Line Segment Tracking is also comparable on the cost front with the CMS track pattern recognition algorithm scaled to 64 cores, since two 32 core Intel Skylake Gold Xeon Processors (commonly used in the multi-CPU efforts) have a similar price to a Tesla V100 GPU. 
	\begin{table}
	    \centering
	    \begin{tabular}{|c|c|}
	        \hline
	        Number of streams & Average time per event (ms) \\
	        \hline
	        1 & 33.7 \\
	        2 & 27.3 \\
	        4 & 26.2 \\
	        6 &  26.3 \\
	        8 & 25.7 \\
	        \hline
	    \end{tabular}
	    \caption{Line Segment Tracking timing performance - \ttbar + PU200, Tesla V100 GPU}
	    \label{tab:timing}
	\end{table}
	
	\section{Conclusions}
    Line Segment Tracking is a parallelizable track reconstruction algorithm targeting the Phase 2 CMS Detector. The backbone of the algorithm is the stub created from the bi-layer module, called the mini-doublet. The higher order objects are successively built by linking lower order objects which can then create entire tracks, with each step of the process being parallelizable. Selections and cuts are applied to reduce the overwhelming combinatorial backgrounds. Good reconstruction efficiency and low fake rates have been achieved on \ttbar + pile-up 200 samples, competitive with the CKF implementation used by CMS. The GPU implementation of the algorithm has innovative memory management and multi-streaming techniques, and provides competitive timing performance on Nvidia Tesla V100 GPUs.
    \paragraph{}
    Our future work will involve revisiting some of the physics selections and looking at methods to reduce the combinatorial background and improve the efficiency even further. A substantial amount of effort will go into improving the reconstruction efficiency of displaced tracks. On the optimization and computational performance front, we have currently only scratched the surface. Our future efforts in this front will go towards efficient computation of mathematical parameters for physics selection, in addition to improving memory throughput by exploring data types such as half precision floats. Plans to improve memory coalescing and optimizing register usage will help in improving our timing performance further. Our final target is to deploy this in the CMS software backend for HLT and offline reconstruction in time for HL-LHC.
    
    \Acknowledgements{This work was supported by the U.S. National Science Foundation under Cooperative Agreements OAC-1836650 and PHY-2121686 and grant NSF-PHY-1912813.}
	\printbibliography
\end{document}

%% file: econfmacros.tex
\def\beq{\begin{equation}}
\def\eeq#1{\label{#1}\end{equation}}
\def\eeqn{\end{equation}}

\def\beqa{\begin{eqnarray}}
\def\eeqa#1{\label{#1}\end{eqnarray}}
\def\eeqan{\end{eqnarray}}

\let\bar=\overbar

\def\Dslash{\not{\hbox{\kern-4pt $D$}}}
\def\dslash{\not{\hbox{\kern-2pt $\del$}}}

\def\msb{{\bar{\ssstyle M \kern -1pt S}}}

